\input epsf                                                               %
\input harvmac
\def\Title#1#2{\rightline{#1}\ifx\answ\bigans\nopagenumbers\pageno0\vskip1in
\else\pageno1\vskip.8in\fi \centerline{\titlefont #2}\vskip .5in}

%
%
\ifx\epsfbox\UnDeFiNeD\message{(NO epsf.tex, FIGURES WILL BE IGNORED)}
\def\figin#1{\vskip2in}
\else\message{(FIGURES WILL BE INCLUDED)}\def\figin#1{#1}
\fi
\def\Fig#1{Fig.~\the\figno\xdef#1{Fig.~\the\figno}\global\advance\figno
 by1}
%
%
%
%
\def\ifig#1#2#3#4{
\goodbreak\midinsert
\figin{\centerline{\epsfysize=#4truein\epsfbox{#3}}}
\narrower\narrower\noindent{\footnotefont
{\bf #1:}  #2\par}
\endinsert
}


\def				
  \Complexes
   {{\rm C}\llap{\vrule height6.3pt width1pt depth-.4pt\phantom t}}

\def\ideq{\equiv}		

\def\=>{\Rightarrow}

\def\semidirect{\mathbin{\hbox{\hskip2pt\vrule height 4.1pt depth -.3pt
                width .25pt \hskip-2pt$\times$}}}
\def\R{{\rm I\!\rm R}}

\def\M{{{}^4\!M}}
\def\f{{{}^4\!f}}

\def\C{{\cal{C}}}
\def\B{{\cal{B}}}
\def\D{{\rm{D}}}

\def\s{{\rm{s}}}

\def\R{\hbox{\rm I \kern-5pt R}}
\def\q{{\tilde q}}
\def\Q{{\widetilde Q}}
\font\ticp=cmcsc10
\def\ajou#1&#2(#3){\ \sl#1\bf#2\rm(19#3)}
\def\sqr#1#2{{\vcenter{\vbox{\hrule height.#2pt
         \hbox{\vrule width.#2pt height#1pt \kern#1pt
            \vrule width.#2pt}
         \hrule height.#2pt}}}}

%

\lref\mackey{G.W.~Mackey, {\it Theory of Unitary Group Representations}
 (University of Chicago Press, 1976).}

\lref\witt{Donald Witt, private communication}  

\lref\unimod{A.~Daughton, J.~Louko and R.D.~Sorkin,
``Initial Conditions and Unitarity in Unimodular Quantum Cosmology'',
    {\it Proceedings of the Fifth Canadian Conference on
      General Relativity and Relativistic Astrophysics} held
           Waterloo, Canada, May, 1993, 
      (World Scientific, to appear)
      (available from the gr-qc electronic archive as Paper: gr-qc/9305016)}

\lref\drexel{
  R.D.~Sorkin,
``Quantum Measure Theory and its Interpretation'', in
    D.H.~Feng and B-L~Hu (eds.),                 
    {\it Proceedings of the Fourth Drexel Symposium on Quantum
       Nonintegrability: Quantum Classical Correspondence},
       held Philadelphia, September 8-11, 1994, pages 205-227
    (International Press, in press),
    $<$e-Print Archive: gr-qc/9507057$>$ }

\lref\cecile{
Michael G.C.~Laidlaw and C\'ecile Morette DeWitt,
``Feynman Functional Integrals for Systems of Indistinguishable Particles'',
 {\it Phys. Rev. D}, {\bf 3}: 1375-1378 (1971)}

\lref\rafsum{R. D. Sorkin and S. Surya,
  ``An Analysis of the Representations of the Mapping Class Group of a 
    Multi-geon Three-manifold''
    (submitted) 
    $<$e-Print archive: gr-qc/9605050$>$}

\lref\rafspinstat{R.D. Sorkin, \ajou Comm. Math. Phys. &115 (88) 421.}
\lref\friedsor{ J.L. Friedman and 
          R. D. Sorkin, \ajou Gen. Rel. Grav. &14 (82) 615.}
\lref\stats{reference on statistics of geons}
\lref\raflect{R.D. Sorkin, ``Classical Topology and Quantum Phases: Quantum
 Geons,'' in {\it{Geometrical and algebraic aspects of nonlinear field
                 theory}}, proceedings of the meeting on geometrical and
                 algebraic 
                 aspects of nonlinear field theory, Amalfi, Italy, May
                 23-28, 1988; eds S. De Filippo, M Marinaro, G. Marmo and G.
                  Vilasi (Amsterdam, New York, North-Holland 1989)}
\lref\rafintro{R.D. Sorkin, ``Introduction to Topological Geons,''
in {\it{Topological properties and global structure of space-time}},
proceedings of the
NATO Advanced Study Institute on Topological Properties and
                 Global Structure of Space-Time, 1985, Erice, Italy;
                         eds P. G. Bergmann and V. De Sabbata.
        (New York, Plenum Press, 1986)}

\lref\raffour{
 R.D.~Sorkin, 
``On the Role of Time in the Sum-over-histories Framework for Gravity'',
    paper presented to the conference on The History of Modern
      Gauge Theories, held Logan, Utah, July, 1987; 
        published in  
          {\it Int. J. Theor. Phys.} {\bf 33}:523-534 (1994) }

\lref\fried{J.L. Friedman and A. Higuchi, \ajou Nucl. Phys. &B339 (90) 491.}
\lref\nico{D. Giulini, ``Properties of Three-Manifolds for
Relativists,'' Freiburg preprint THEP- 93/15, gr-qc/9308008}
\lref\hartlewitt{J.B. Hartle and D.M. Witt, \ajou Phys. Rev. &D37 (88) 2833.}
\lref\milnor{J. Milnor, ``Introduction to algebraic K-theory,'' Annals of
                 Mathematics Studies Vol. 72, (Princeton University Press,
                 1971).}   
\lref\giulou{D. Giulini and J. Louko, \ajou Phys. Rev. &D46 (92)
                 4355.}
\lref\friedwitt{J.L. Friedman and D.M. Witt, \ajou Topology &25 (86) 35.}


\Title{\vbox{\baselineskip12pt
\hbox{CALT-68-2068}
\hbox{qr-qc/9609064}}
}
{\vbox{\centerline {A Spin-Statistics Theorem}\vskip2pt
\centerline{for Certain Topological Geons}
}}

\centerline{{\ticp H.F. Dowker}${}^{1}$}
\centerline{{\ticp R.D. Sorkin}${}^{2}$}

\vskip.1in

\centerline{\sl ${}^1$Lauritsen Laboratory}
\centerline{\sl California Institute of Technology}
\centerline{\sl Pasadena CA 91125, U.S.A.}

\vskip.1in

\centerline {\sl ${}^2$Instituto de Ciencias Nucleares}
\centerline {\sl UNAM, A. Postal 70-543}
\centerline {\sl D.F. 04510, Mexico}
\centerline{and}
\centerline{\sl Physics Department}
\centerline{\sl Syracuse University }
\centerline{\sl Syracuse, NY 13244-1130, U.S.A.}

\bigskip

\centerline{\bf Abstract}
We review the mechanism in quantum gravity whereby topological geons,
particles made from non-trivial spatial topology, are endowed with
nontrivial spin and statistics.  In a theory without topology change there
is no obstruction to ``anomalous'' spin-statistics pairings for geons.
However, in a sum-over-histories formulation including topology change, we
show that non-chiral abelian geons do satisfy a spin-statistics correlation
if they are described by a wave function which is given by a functional
integral over metrics on a particular four-manifold.  This manifold
describes a topology changing process which creates a pair of geons from
$\R^3$.

\Date{September 1996}

\newsec{Introduction}

When the configuration space of a classical system is non-simply-connected
(or, more generally, non-contractible) it allows for a richer variety of
possibilities quantum mechanically than usual. In particular, the
possibility arises of ``emergent'' fermionic statistics and spinorial
(half-odd-integral) spin for objects built from fields which are
fundamentally tensorial (integral spin) and bosonic.  General relativity is
such a classical theory and in quantum general relativity on a product
spacetime manifold $\R\times\,^3\!M$ it can be shown that topological geons may
be endowed with both nontrivial spin \friedsor\ and nontrivial statistics
\rafintro.

In nature, the spin and statistics of all known particles
are correlated: they are bosons if and only if they are tensorial,
and fermions if and only if they are spinorial.  Quantum geons, on the
other hand, satisfy no such correlation in the canonical theory: any
combination of spin and statistics is possible \refs{\rafintro,
\raflect, \rafsum}. It is perhaps not surprising that geons in canonical
quantum gravity appear to violate the usual correlation. In the proofs of
all existing spin-statistics theorems, the (explicit or implicit)
possibility of particle-anti-particle creation (and annihilation) is
crucial.  But the process of geon-anti-geon pair production is a topology
changing one and cannot be described within a formalism which assumes, a
priori, that the spatial three-manifold is fixed. It has therefore been
conjectured  that in a formulation of quantum gravity which can
accommodate topology change, the usual spin-statistics connection would be
recovered for geons \raflect.

One such formulation is the sum-over-histories (SOH) with inclusion of
non-product spacetime manifolds.  In this paper we will show that for
certain geons, a spin-statistics theorem can be proved in this context, one
which relies on the wave function describing the quantum state of the geons
being given by a  functional integral on a certain ``U-tube'' manifold.

In section 2 we will briefly describe the way in which topological geons
acquire spin and statistics.  In section 3 we adopt a SOH 
approach to quantum gravity and introduce the key assumption that the geons
we consider are described by a functional integral over metrics on a
certain four-manifold, $\M$, that mediates the pair production of the
geons.  We will see how this implies that a certain diffeomorphism
intimately connected with spin and statistics, acts trivially on the
wave function.  In section 4 we show that for a particular sort of geon, the
lens spaces, this result leads to a spin statistics theorem: the lens space
geons (which are necessarily tensorial) must always be bosons.  More
generally, the theorem we prove applies to any geon which carries an {\it
abelian} representation of its internal diffeomorphism group.  Section 5 is
a summary and discussion of possible extensions of this work.

 In order to avoid repeatedly having to make certain caveats, we will
restrict ourselves to orientable three-manifolds in this paper; this
restriction could straightforwardly be dropped and we
will mention in the final section how our results generalize 
to the non-orientable case.  
We will further assume
that no handles, $S^2 \times S^1$, occur in the ``prime decomposition'' of
the three manifold (see section 2).

\newsec{Topological Geons}

In this section we sketchily review the background to our problem,
referring the reader to \refs{\rafintro,\raflect,\rafsum} for more details.
Roughly, topological geons are particles made from non-trivial spatial
topology.  We will be interested in the situation of an isolated system of
particles and thus we will be dealing with a three-dimensional manifold $M$
which admits asymptotically flat metrics.  Physically, $M$ is three-space
at a ``moment of time'', or if you prefer, the ``future boundary of
truncated spacetime'' \refs{\raffour}.

\subsec{The Topology}

There is a ``Three-Manifold Decomposition Theorem'' that identifies
candidates for elementary geons, but in order to state this theorem we must
first introduce the concepts of ``connected sum'' and ``prime manifold.''
To take the connected sum of two oriented three-manifolds $M_1$ and $M_2$,
remove an open ball from each and identify the resulting two-sphere
boundaries with an orientation-reversing diffeomorphism (henceforth,
``diffeo'').  Taking the connected sum of any three-manifold with $S^3$
gives a manifold diffeomorphic to the original one; taking it with $\R^3$
is topologically equivalent to deleting a point.  A prime three-manifold,
$P$, is a closed three-manifold that is not $S^3$ and such that whenever
$P=P_1\# P_2$, either $P_1$ or $P_2$ is $S^3$. Examples of primes are the
three-torus, $T^3$, and the so-called spherical spaces, $S^3/G$, where $G$
is some discrete subgroup of $SO(4)$ acting freely on $S^3$.

The $M$ we are considering is $M=\R^3 \# K$ where K is a closed
three-manifold.  The Decomposition Theorem states that any such $M$ can be
decomposed into the connected sum of finitely many prime manifolds and this
decomposition is unique:
   \eqn\decomp{M=\R^3 \# P_1 \# P_2 \dots \# P_n.}
We will assume that to each prime summand there corresponds an elementary
quantum geon; with ``correspond'' being used in a suitable sense since
there is a rather subtle relation between a particular piece of spatial
topology and a physical particle---which subtlety has to do both with
familiar ``identical particle exchange effects'' and unfamiliar effects due
to the existence of diffeos known as ``slides'' \refs{\raflect, \rafsum}.

A very useful way to visualize a multi-geon manifold relies on the result
that any prime manifold (in fact any closed three-manifold) can be
constructed from a solid convex polyhedron by performing appropriate
identifications on its faces. For example, the three-torus prime, $T^3$, is
made by identifying opposite faces of a solid cube. The spherical space
$S^3/Q$, where Q is the eight-element quaternion subgroup of $SU(2)$, is
also made from a solid cube, this time identifying opposite faces after a
${1\over 2}\pi$ rotation. Suppose $M=\R^3\# P$ where $P$ is a prime
expressed as a certain solid polyhedron with identifications. Then $M$ is
diffeomorphic to $ P\backslash{\rm\{point\}}$. By letting the point be
removed from the interior of $P$ and imagining ``turning $P$ inside out,''
one sees that $M$ can be constructed by {\it{deleting}} the same (open)
solid polyhedron from $\R^3$ and making the same identifications on the
boundary. In the same way the multi-geon manifold $M=\R^3 \# P_1 \# P_2 \#
\dots \# P_n$ can be made by cutting out an appropriate polyhedron 
 from $\R^3$ for each summand  and making appropriate identifications. 

\subsec{Wave Functions and the Mapping Class Group}

In canonical quantum gravity, for which the topology does not change, the
configuration space, $Q$, is the space of all three-geometries on $M$,
 \eqn\config{Q= {{{\rm{Riem}}^\infty(M)}\over{{\rm{Diff}}^\infty(M)}}}
where ${{\rm{Riem}}^\infty(M)}$ (${\rm{R}}^\infty$ for short) is the space
of asymptotically flat Riemannian metrics on $M$ and
${{\rm{Diff}}^\infty(M)}$ ($\D^\infty$ for short) is the group of
diffeomorphisms of $M$ that become trivial on approach to
infinity.\foot
 {By asymptotically flat we mean asymptotic to some fixed, flat metric in a
 neighborhood of infinity.  The exact space $Q$ will depend on the precise
 falloff conditions imposed on the metrics and diffeomorphisms, but we will
 only be interested here in topological properties of all the spaces which
 are insensitive to the choice of conditions.}
It can be shown that ${\D}^\infty$ acts freely on ${\rm{R}}^\infty$ and so
$Q$ is a manifold, $R^\infty$ being a principal fibre bundle over $Q$ with
fibre $D^\infty$.  Thus, using the fact that $R^\infty$ is convex and hence
contractible to a point so that all its homotopy groups are trivial, we
deduce that $\pi_k(D^\infty)\simeq\pi_{k+1}(Q)$.

Wave functions need not be single-valued on $Q$ if $Q$ contains
non-contractible loops.  Rather, the transformation of a wave function as
such loops are traversed gives a representation of $\pi_1(Q)$.  This is a
special case of the general situation where wave functions are sections of
a twisted vector bundle on $Q$.

In the so-called covering space quantization, wave functions can be
represented as (single-valued) complex functions on the universal covering
space of $Q$,
\eqn\cover
  {\widetilde Q ={{{\rm{Riem}}^\infty(M)}\over{{\rm{Diff}}_0^\infty(M)}}}
where ${\rm{Diff}}_0^\infty(M)\subset {{\rm{Diff}}^\infty(M)} $ is the
connected component of the identity.  The space $\Q$ is a principal fibre
bundle over $Q$ with fibre $G=\pi_0(\D^\infty) := \D^\infty/\D_0^\infty$
\eqn\princ{\matrix {G&{\buildrel i \over \longrightarrow}&\Q\cr
           {}&{}&\;\;\bigg\downarrow \pi\cr
           {}&{}& Q\cr}}
The group $G$, known as the mapping class group (MCG) of $M$,
acts (globally) on the right on $\Q$: if $g\in G$
and $\tilde q\in \Q$ then, under the action of $g$, $\tilde q \mapsto \q g$.
This action can be given in terms of metrics and diffeomorphisms as follows.
Let $h$ be some representative metric of the equivalence class
$\q$, $\q=[h]$, and $d$ some representative diffeomorphism of the equivalence
class $g$. Then 
 \eqn\pull{ \q g = \left[ d^{*} (h) \right] }
where $d^{*}(h)$ is the pullback of $h$ by $d$.  The action induces an
action on functions on $\Q$ by the requirement that they transform as
scalars. More specifically, under $g$, ${\Psi\mapsto \Psi g }$ where 
  \eqn\diff{ \left( \Psi g \right) (\tilde q) =\Psi (\q g^{-1}).}
Notice our convention of writing the action of $g$ on the right of $\Psi$
to agree with its action on $\Q$.

If, now, we consider square-summable functions $\psi$ on a single fibre of
the covering space $\Q$, then \diff\ says precisely that the space of such
functions carries the {\it regular representation} $R$ of $G$.  When $G$ is
a finite group, every unitary irreducible representation of G occurs as a
subrepresentation of $R$, and conversely $R$ can be decomposed uniquely as
a sum of irreducibles.  Physically each such irreducible corresponds to a
distinct sector (``theta sector'') of the quantum theory, and these sectors
will be superselected if topology change is ignored.  (Note that, since an
irreducible of dimension $d$ occurs $d$ times in $R$, there will be more
irreducible subspaces than physically distinct sectors.  See \refs{\rafintro}
on this point.)  When $G$ is infinite (which the MCG almost always is), these
statements about irreducibles must be replaced by a considerably more
complicated set of assertions \refs{\mackey}. In part, this is just the
familiar problem of delta-function normalization for eigenvectors
of operators with a continuous spectrum, but that is not the whole story.
Nevertheless, we believe that it remains formally true that every
irreducible can, in an appropriate sense, be obtained from the regular
representation, at least if we limit ourselves to irreducibles of finite
dimension.  We return briefly to this point near the end of the paper,
where we argue that the difficulty belongs to the wave function rather than
the physics, and will disappear if one adheres consistently to a
sum-over-histories formulation.

Equation \diff\ defines
an action of $G$ on the space of wave functions on $\Q$. For
consideration of spin and statistics properties we are actually interested
in the action of {\it loops} in $Q$ (elements of $\pi_1(Q)$) on wave
functions. Although $G$ is isomorphic to $\pi_1(Q)$ the isomorphism is not
canonical. Another way to say this is that diffeos act (globally) on the
right on the bundle $\Q$, whereas loops act (locally) on the left.  Let us
see what this means in more detail by constructing an isomorphism from
$\pi_1(Q)$ to $G$.

More specifically, let $\pi_1(Q)=\pi_1(Q;q_0)$ be the first homotopy group
{\it based at} $q_0\in{}Q$, and let $\gamma$ be a representative of the
homotopy class $[\gamma]\in\pi_1(Q;q_0)$.  Now
$[\gamma]$ induces an automorphism of $\pi^{-1}(q_0)$, the fibre above $q_0$,
by sending the point $\q\in\pi^{-1}(q_0)$ to the point at the end of the
path in $\Q$ which is the unique lift of $\gamma$ that begins at
$\q$.  Choose a fiducial element, ${\q}_0 \in \pi^{-1}(q_0)$, then
$\pi^{-1}(q_0) = \{{\q}_0 f: f\in G\}$, and the requirement
  \eqn\lift{ [\gamma]: \q_0 \mapsto \q_0 g_{\gamma}}
sets up a correspondence $\Phi$ between $[\gamma]\in\pi_1(Q)$ and
$g_{\gamma}\in{}G$ which depends on the choice of $\q_0$ (but not on the
choice of $\gamma$ to represent $[\gamma]$).  Now the action of the loop
$\gamma$ on a general point on the fibre, $\q = \q_0 f_{\q}$ with $f_{\q}
\in G$, is
  \eqn\loopact{ 
  [\gamma]: \q =\q_0 f_{\q}\mapsto 
  [\gamma](\q_0 f_\q) =
  \left( [\gamma] \q_0 \right) f_\q =
  \q_0 g_{\gamma} f_{\q} \,
  \left[ = \q (f_\q^{-1} g_\gamma f_\q) \right] }
from which we can deduce that ($i$) $\Phi: [\gamma] \mapsto g_\gamma$ is an
isomorphism and ($ii$) the isomorphism is not canonical since choosing a
different fiducial point on the fibre gives a different isomorphism
(related by conjugation).

Let $\Psi$ be a wave function on $\Q$ with support on the fibre
$\pi^{-1}(q_0)$.  The action via {\loopact} of $\pi_1(Q)$ on this fibre
induces an action on $\Psi$, $\Psi \mapsto [\gamma] \Psi$ where
 \eqn\induce{
 [\gamma]\Psi(\q) = \Psi(\q_0 g_{\gamma}^{-1} f_{\q})\quad.}
(Compare this to \diff.)  We can extend all of this to a neighbourhood of
the fibre in which the bundle is a product, since a loop through $q_0$ maps
unambiguously to a loop at a neighbouring point $q_0'$ within such a
neighbourhood.  Notice that the loop $[\gamma]$ acts trivially on a
wave function $\Psi$ (i.e. leaves it invariant) if the diffeo $g_{\gamma}$
and all conjugates of it, $g g_\gamma g^{-1}$, act trivially on $\Psi$.


As we have already done, we will often refer, imprecisely, to an element of
$\pi_0(\D^\infty)$ as a diffeo and an element of $\pi_1(Q)$ as a loop but
there should be no ambiguity involved. We will actually never be interested
in loops and diffeos as such but only in their homotopy and isotopy
classes.

\subsec{Spin and Statistics}

Now let $M = \R^3 \# P \# P$ and let $q \in Q$ be a configuration in which
two isometric geons are sitting at well-separated positions with plenty of
flat space between them.  Call the loop based at $q$ which describes the
two geons moving (by translation) around each other till they have swapped
places the ``exchange loop,'' $\gamma_e$, and the loop that describes one
geon spinning around by $2\pi$ the ``$2\pi$ rotation loop'' of that geon,
$\gamma_{2\pi}^i$, where $i=1,2$ labels the geon (by its physical
position).  Further, suppose we have a wave function, $\Psi$, on $\Q$ which
is peaked on the fibre over $q$.

%
%

If the $2\pi$ rotation loop of one geon is represented on $\Psi$ by 1
$(-1)$ then that geon is tensorial (spinorial). If the exchange is
represented by 1 ($-1$) and the two geons are in identical internal states
then the geons are bosons (fermions). In a version of quantum gravity in
which the three-manifold M is fixed, there is no correlation between the
spin type and statistics that geons can have. In the case of two identical
primes, there exist finite dimensional
unitary irreducible representations of the 
MCG for each of the possible combinations: fermion-tensorial,
fermion-spinorial, boson-tensorial and boson-spinorial
\refs{\rafsum}.
 

This lack of a correlation can be attributed to the fact that
in a frozen topology  theory (such as canonical quantization)
 there is no
allowance for geon-anti-geon production since
a process in which a geon
and anti-geon are created from $\R^3$ is a topology changing one (we know this
from the decomposition theorem: one piece of non-trivial topology can't
``cancel'' another). The known spin-statistics theorems
for objects such as skyrmions and other kinks which have these ``emergent'' 
properties of spin and statistics all require, for their proofs, that the
process of pair creation and annihilation be describable as a path in
the configuration space.
For two $SU(2)$ skyrmions for example, the exchange
loop and the $2\pi$ rotation loops in the two-skyrmion
sector of the configuration space can be shown to be homotopic,
and therefore must be represented on the state vector identically. 
The homotopy
sequence of loops leading from the exchange to one of the $2\pi$ rotations
contains a loop which describes a skyrmion-anti-skyrmion pair emerging 
from the vacuum and the anti-skyrmion annihilating with one of the 
original skyrmions to leave two skyrmions again. See also \refs{\rafspinstat}
for a more general theorem. 

All this leads one to expect that in a formulation of quantum gravity in
which topology change is naturally accommodated there is hope that the spin
statistics correlation can be recovered.\foot
{One could take the position that the unreasonably large number of
 inequivalent quantum sectors which arise in the canonical theory, due to
 the effect of the slides discussed in subsection 3.3, is another reason to
 abandon the assumption of frozen topology \refs{\rafsum}.  }
We will therefore turn now to the SOH approach.

\newsec{Sum-Over-Histories}

\subsec{The Wavefunction}

Henceforth we take $M= \R^3 \# P \# P$ where $P$ is a non-chiral prime,
that is, a prime which admits an orientation
reversing diffeo.  Construct a four-manifold $\M$ with ``initial''
boundary $M_0 =\R^3$ and ``final'' boundary $M = \R^3 \# P \# P$ by taking
$\R^3 \times I$, where I is the unit interval and deleting a ``U-tube'' of
polyhedral cross-section. Figure 1 is a depiction of this in 2+1 dimensions
-- the generalization to 3+1 should be clear.

\ifig{\Fig\utube}{$\M$}{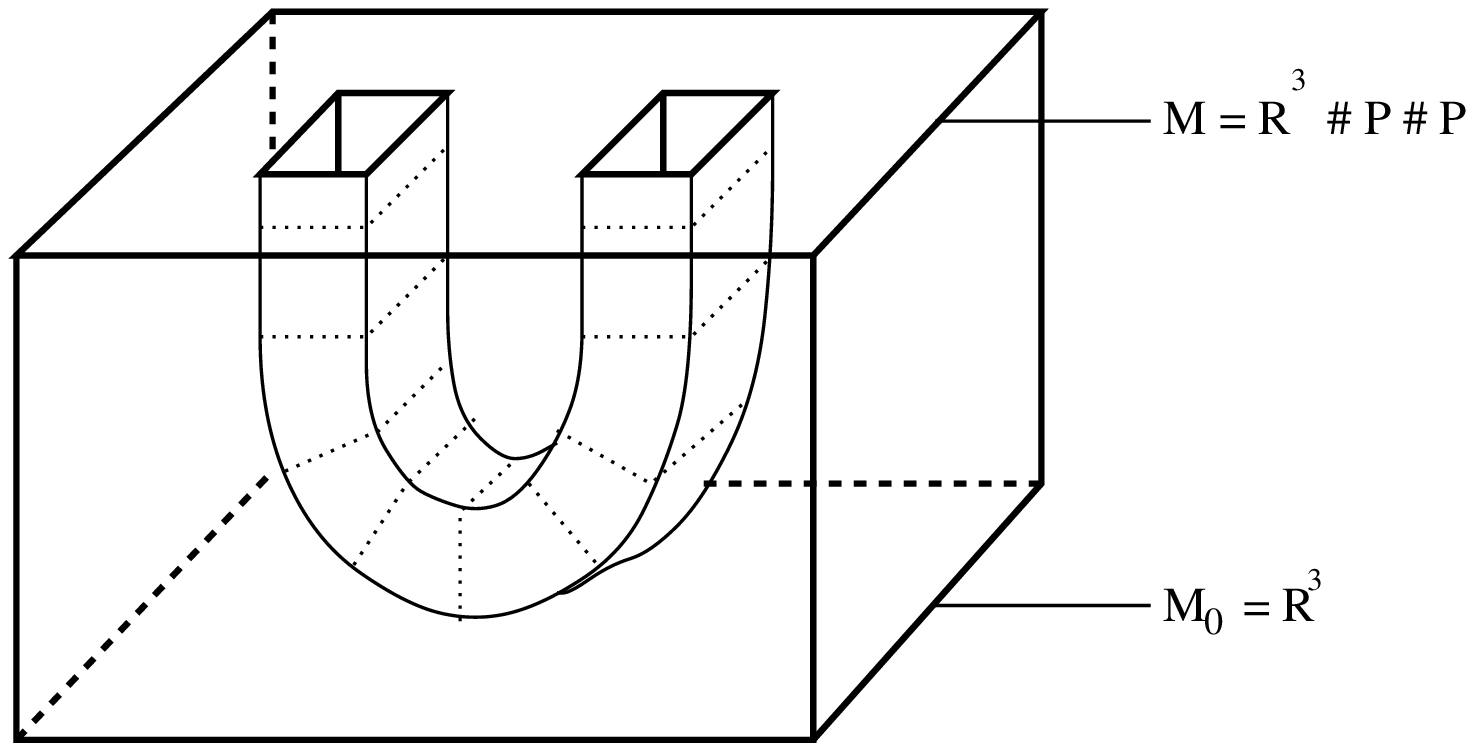}{2.2}

The tube is drawn with square (imagine cubical) cross-section, appropriate
for torus geons for example; in general the cross-section will be a more
complicated polyhedron.  Identifications are made on each cross-section
(shown with dotted lines) of the cut-out tube's boundary, just as in
constructing $\R^3 \# P$.  The condition that $P$ be non-chiral is necessary
for the existence of this four-manifold: if one end of the tube had
identifications made on it that made it a chiral prime, $\hat P$, then the
other end would be a different prime, ${\hat P}^{*}$, its ``$CP$
conjugate.''

Consider wave functions given by a functional integral of the following 
form:
 \eqn\wave{\Psi(h) = \int_{\B} [dh_0] \Psi_0(h_0) \int_{C} [dg] e^{i S[g]}}
where $\B$ is the class of all asymptotically flat three-metrics on $M_0$,
$\Psi_0$ is any wave function on $\B$, $\C$ is the class of all
four-metrics on the four-manifold $\M$ which induce $h_0$ on $M_0$ and $h$
on $M$ and approach some fixed flat metric at infinity. 
The class $\C$ could be more restricted: one might want to sum over
metrics with a fixed four-volume for example \raffour.
%
%
Note that $\Psi$ is given as a function on ${\rm{R}}^\infty$, whereas we
want it to be a function on $\Q$.  In fact it defines a function on $\Q$
since it is constant on equivalence classes of metrics related by diffeos
$j$
connected to the identity.

This follows from the more general result that any
element, $f\in\D^\infty$ that admits an extension $\f:\M\rightarrow\M$
which is the identity on $M_0$, and which tends to the identity at infinity,
acts trivially on \wave.  [From now on, we will take for granted that any
diffeo of $^4\!M$ which we consider must tend to the identity at infinity.
Also notice that since $\pi_0\left(\D^\infty(\R^3)\right)$ is trivial the
question of whether there exists an extension which fixes $M_0$ pointwise
reduces to the question of whether there exists any extension at all.]
This holds if
the ``measure-factor''  $[dg]$ and amplitude $e^{iS[g]}$ in \wave\ are
diffeomorphism invariant. Indeed, consider $\Psi(h)$ 
and $\left(\Psi f\right) (h) = \Psi ( f^{*}(h))$. If $\f$ exists such that
its restriction to the initial boundary is the identity and to the final
boundary is $f$, then for each metric $g$ contributing to $\Psi$ there is
a diffeomorphic partner, $\f^*(g)$ contributing the same amount to $\Psi f$
and vice versa. Thus the two wave functions are equal.  

Now, any diffeo,
$j\in\D^\infty$ connected to the identity is extendible to a diffeo of $\M$
\hartlewitt.  Briefly, this can be seen by constructing an extension which
is the identity outside a neighbourhood of the boundary $M$; on the
neighbourhood, which is diffeomorphic to $M\times[0,1]$, it is defined
using the isotopy sequence of diffeos between $j$ and the identity.  The
wave function constructed in \wave\ is therefore invariant under $j$.

We should also confirm that extendibility is an isotopy invariant
property, 
{\it i.e.} if $g,g'\in[g]$ and $g$ is extendible to $\M$, then so is
$g'$.  Let the extension of $g$ be $^4\!g:\M\rightarrow\M$ and let the
isotopy sequence be $i:[0,1]\rightarrow \D^\infty(M)$ with $i(0)=g'$ and
$i(1)=g$. Consider an open neighbourhood, $N$, of $M$ in $\M$. There exist
asymptotically trivial diffeos: $\chi: N \rightarrow [0,1) \times M$ and
$\theta: \M \backslash N\rightarrow \M$.  We define $^4\!g'$ as follows.
For $x\in N$, $(x)\chi = (s, y)$ with $s\in [0,1)$ and $y\in M$, and we set
$(x)^4\!g' = (y)(i(s))$. For $x\in\M \backslash N$ we set $(x)^4\!g' = (x)
\theta ^4\!g$. Then $^4\!g'$ extends $g'$ to $\M$.

\subsec{A Special Diffeomorphism $F$}

We are interested in the action on $\Psi$ of the diffeos
$f_e \equiv g_{\gamma_e}$ and $ f_{2\pi} \equiv 
 g_{\gamma^1_{2\pi}}$ (see section 2) of $M$.
We recall here
the notion of the ``development'' of a diffeo by a sequence of manifolds. 
For more details see \refs{\rafintro, \raflect}. 
$M$ is constructed by taking $\R^3$ and 
cutting out two polyhedra and making identifications. We construct a  
continuous sequence
of such manifolds, cutting out the polyhedra in slightly different positions
each time, the final one being $M$ again, the sequence thus being a ``loop
of manifolds.''\foot
{That this
sequence of manifolds is continuous in some appropriate sense 
seems clear, though this statement cannot as yet 
be given a precise meaning since
no topology on the space of manifolds has been exhibited with respect 
to which continuity could be defined.}
 Each manifold in the sequence, 
$M(s)$,
is diffeomorphic to $M$, so there exists a continuous sequence of diffeos
$f(s): M \rightarrow M(s)$, $s \in [0,1]$. The final diffeo $f(1)$ is 
a diffeo from $M$ to itself and we say that it is developed by the loop
$M(s)$. Every diffeo is developed by some loop and two diffeos developed by the  same loop are isotopic (i.e there exists a continuous sequence of diffeos
that interpolates between them) as are diffeos developed by 
homotopic loops of manifolds.

Suppose we choose, as fiducial point, $\q_0$, on the fibre, 
a metric which is flat outside two two-spheres, each surrounding one
of the
cut-out polyhedra, such that the isometry between the metrics 
inside the spheres is realised by translation through the flat region. 
Then the diffeo $f_e$ is developed by the sequence of manifolds which 
begins and ends with $M$ and in which the polyhedra are cut out at positions
which move gradually around each other (with fixed orientation with respect
to infinity) until they have swapped places. $f_{2\pi}^i$, $i = 1,2$, is 
developed  by the sequence of manifolds, beginning and ending with $M$, in 
which one polyhedron is cut out at gradually rotated positions until 
it has rotated a whole turn and the other is cut out in the same fixed position all
the time. It is clear that these diffeos represent physical exchange and 
$2\pi$-rotation for the metric $\q_0$.  
 Let $F=f_e f_{2\pi}^1$.

We saw that a diffeo acts trivially on $\Psi$ if it is 
extendible to $\M$. We now show that $F$
is extendible. To do so we exhibit the sequence of four-manifolds
which develops the extension. It starts with $\M$, and in the sequence 
the ends of the cut out tube swap positions, and then one end rotates around
by $2\pi$. The cut out tube itself gets twisted and then untwisted in 
the process so that the final manifold is $\M$ again. Figure 2 is a
2+1 depiction of the sequence. 

\ifig{\Fig\sequence}
{These are nine snapshots of the sequence of manifolds that develops the
extension of $F$ to $\M$ (here represented as 3-dimensional). Only the
cut-out tube is drawn in pictures 2-9, the surrounding manifold is
implied.}{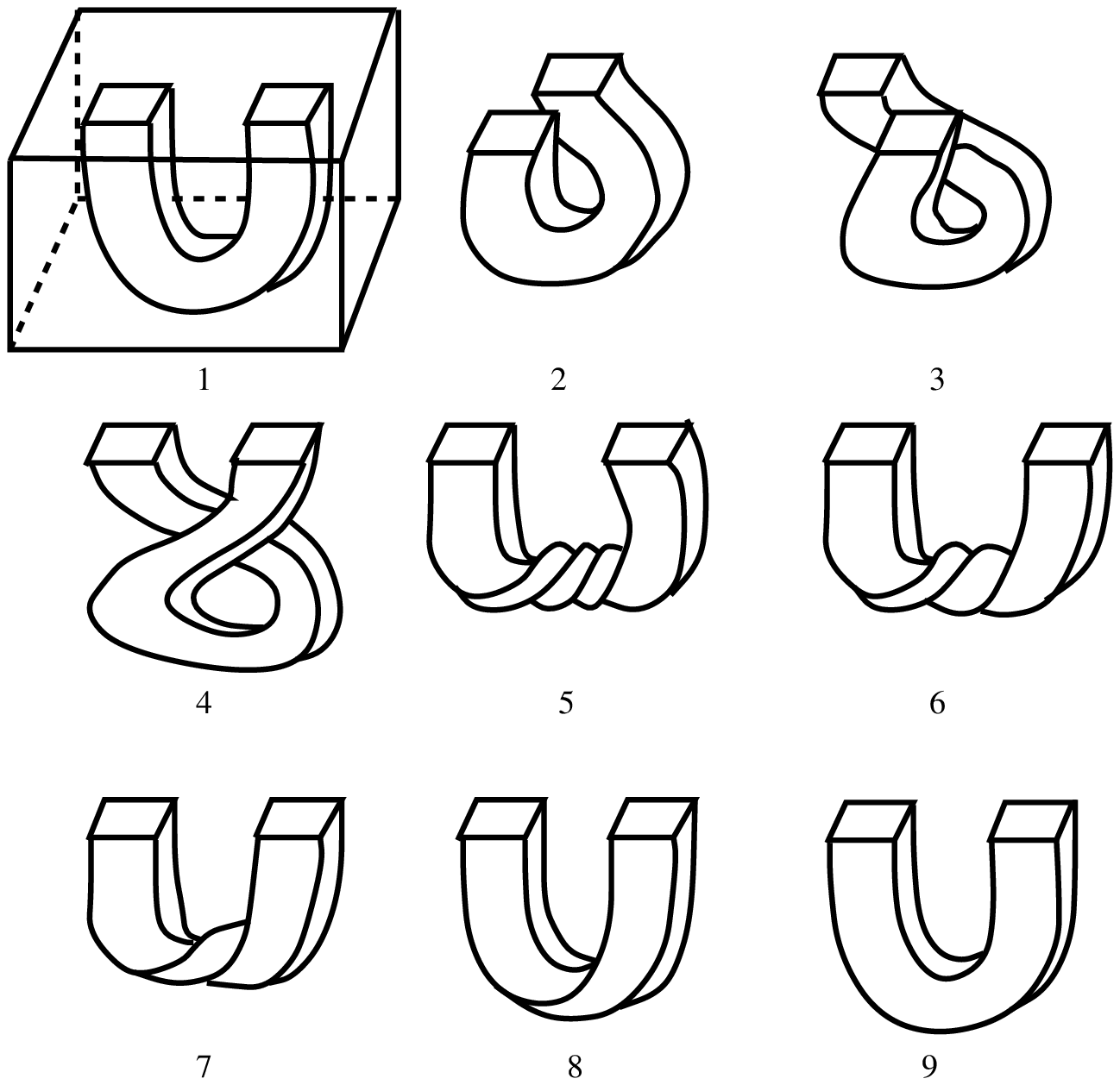}{4}

To see that it also works in 3+1, let's first 
re-express the 2+1 pictures in terms of ``framed curves.''
We regard the pictures in figure 2 as manifolds induced
from framed curves in the $t < 0$ portion of $\R^3$.  The curve itself gives
the location of the tube, and the framing tell how it ``twists".  A
framing just means attaching to each point  of the curve a pair of
labeled unit vectors orthogonal to the curve, and  one can 
cut out the polyhedron and make identifications appropriate 
to  a given prime at each point of
the framed curve in a
canonical manner.

Now, $F$ (still in $2+1$ dimensions)
is developed by a sequence of manifolds,
which under our correspondence, would be a loop in the space of pairs of
``framed points'' (i.e. just frames) in $\R^2$. The framed points swap places 
without rotating and then one of them rotates by $2\pi$.  What figure 2
effectively shows is 
how to extend this sequence of frame-pairs to a sequence of framed curves.
But this gives us a sequence of 3-manifolds, whose boundaries develop
$F$ (a diffeo of $M$); hence the diffeo of $M$ that they develop
extends $F$ from $M$ to $^3\!M$.

So far we have just reformulated the 2+1 proof.  The generalization to any
higher dimension is simple: just regard $\R^3$ as a subspace of $\R^{n-1}$
and complete the 2-frames to $(n-2)$-frames by adding a constant
$(n-4)$-frame in the orthogonal directions.  Thus the
exchange-cum-$2\pi$-rotation of the two frames in $\R^{n-1}$ extends to a
continuous deformation of the framed U-curve in $\R^n$.  As before we can
glue a fixed (non-chiral) geon onto the framed curves to turn the loop of
framed curves into a loop of manifolds which develops the extension of $F$
to $^n\!M$.

We note that figure 2 is slightly misleading in that it actually matters in
2+1 dimensions which way the end of the tube is rotated: one way the tube
untwists, the other way the tube becomes more twisted. In 3+1 this is not
the case since a $4\pi$-rotation is connected to the identity.

We have now shown that the diffeo $F$ acts trivially on any $\Psi$ of 
the form \wave. 

\subsec{The Mapping Class Group}

Although it is clear that the diffeo $F$ is intimately connected with spin
and statistics, we are still some way from a spin-statistics theorem.  For
one thing, the physical 
exchange-cum-$2\pi$-rotation
 loop $\Gamma=\gamma_e\gamma^1_{2\pi}$ only
acts on a wave function peaked around a two geon configuration, $q$, as
described in section 2.2.  Even then, the loop $\Gamma$ only corresponds to
the diffeo $F$ for a particular representative metric of $q$.  To proceed,
we will require more information about how other elements of the mapping
class group, $G$, act on wave functions such as \wave.

Relative to a presentation of $M$ as a connected sum, $G$ is generated by
three sorts of elements (see for example \refs{\raflect}):
(i) the generators of the internal diffeos of one of the primes,
(ii) the exchange diffeo, and
(iii) the slide of one prime through the other.
In fact $G$ takes the form $G\simeq(S\semidirect G_{\rm int})\semidirect E$, 
where $S$ is the normal subgroup generated by the slide, $G_{\rm int}$ is
the internal group, $E\simeq Z_2$ is the subgroup generated by the exchange
alone and $\semidirect$ denotes semidirect product \refs{\rafsum}.
Further, $G_{\rm int} = G_1 \times G_2$, where $G_1$ and $G_2$ are the
internal groups of each separate geon and are isomorphic, which isomorphism
is realised by translation, due to the particular presentation we have
chosen.

A slide can be visualized by imagining one prime shrunk down to a tiny size
and moved around some fixed non-contractible loop through the other one so
that the resultant diffeo is the identity in the interior of some
two-sphere surrounding the geon doing the sliding.  There is a two-fold
ambiguity in this definition of the slide which comes from the fact that
the prime that's being slid can undergo a $2\pi$-rotation while it is on
its journey.  (This ambiguity can be removed by specifying that the
orientation of the slid prime
be fixed 
with respect to some background field of frames.)

Suppose $[s]$ is the isotopy class of the slide diffeo. Consider
$s\in[s]$ such that $s$ is the identity outside some embedded $S^2$ in $M$
surrounding the two primes. Extend the $S^2$ to an embedded cylinder
$S^2\times[0,1]$ in $\M$ as shown in figure 3 and consider $\M'$, the
compact manifold with boundary, formed by cutting off $\M$ outside the
cylinder. Let $M' = \partial \M'$ so that $M' \simeq P\# P$.

\ifig{\Fig\compact}{$\M'$. Strictly speaking, for $\M'$ to be a 
differentiable
manifold, its ``edges'' must be smoothed.}{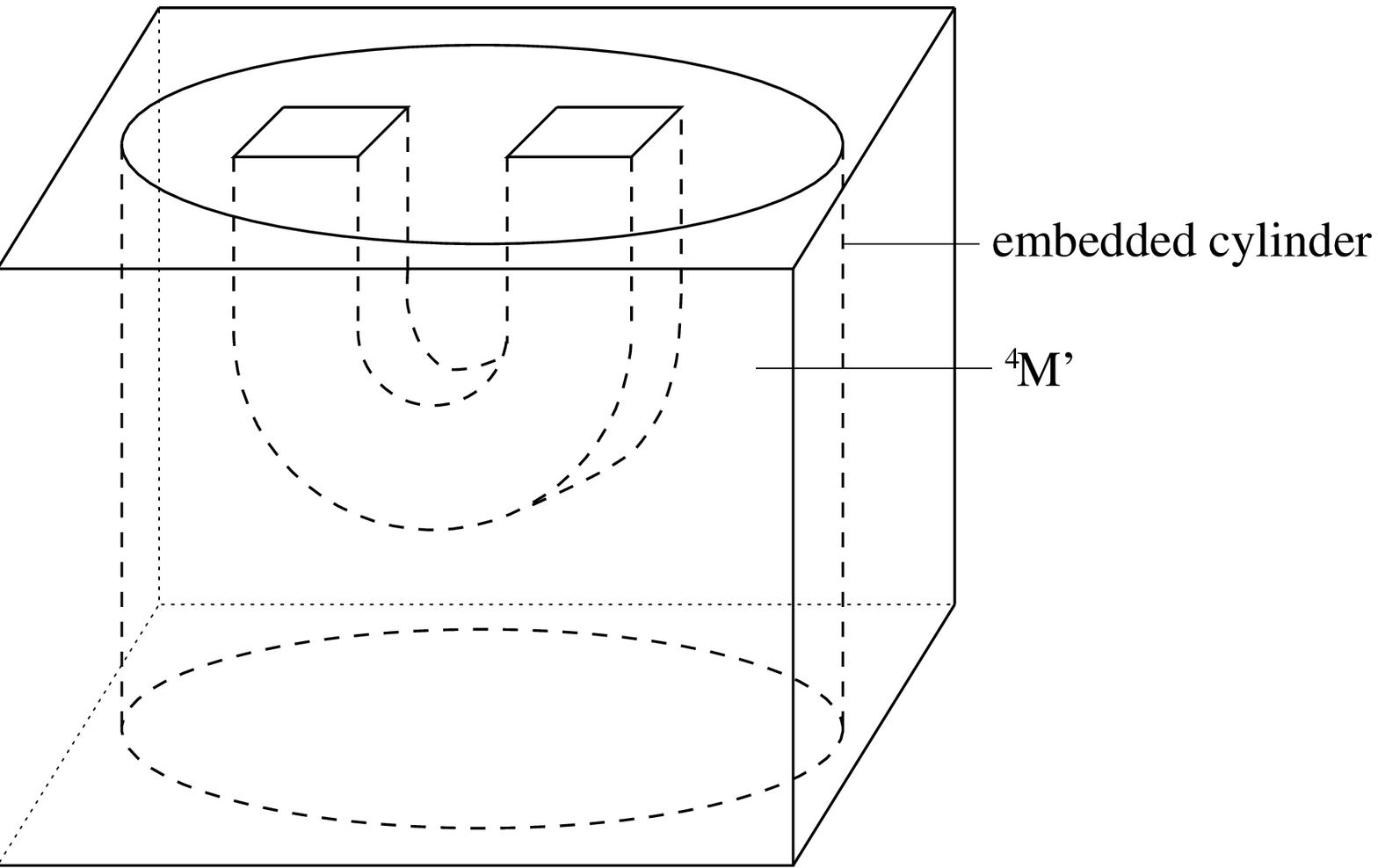}{2.2}

Then $s$ can be extended trivially to an element of $\D_B(M')$, the group
of diffeos of $M'$ which fix the bottom and sides of $M'$.  This extended
$s$ can also be regarded as an element of $\D(M')$, the group of diffeos of
$M'$.  We claim that, within $\D(M')$, $s$ is in fact isotopic either to
the identity or to the $2\pi$-rotation of one of the primes (this being the
two-fold ambiguity in the definition of the slide mentioned above).  The
reason is that the two-sphere separating the two primes in $M'$ is unique
(up to homotopy).  The slide is the identity inside a two-sphere
surrounding one prime, which we can take to be the separating two-sphere.
The slide must therefore be isotopic to an internal diffeo of the other
prime.  A slide, however, has a characteristic action on the generators of
the fundamental group of $M'$.  In particular it leaves the generators
which thread the ``prime through which the other is slid" invariant.  The
only internal diffeo which does this is the $2\pi$-rotation.  Hence the
slide is isotopic either to the identity or to the $2\pi$-rotation in
$\D(M')$.  We assume that we have chosen the slide to be the one isotopic
to the identity in $\D(M')$.\foot
 {We thank Bob Gompf of the University of Texas at Austin for pointing out
 the triviality of the slide to us and providing this argument.  Strictly
 speaking, we should also check that the $2\pi$ rotation $R$ of a single
 prime does not belong to the slide subgroup $S$, when the rotation
 ambiguity in the slide is resolved as above.  But: slide trivial in
 $D(M')$ $\=>$ $S$ trivial in $D(M')$, whereas $R$ is {\it not} trivial in
 $D(M')$ (for spinorial $P$); hence $R\notin S$.  Finally, we note that
 the triviality of the slide in $D(M')$ can also be established directly
 by constructing an explicit deformation of it to the identity.}

Now, any diffeo $g\in \D_B(M')$ which is isotopic to the identity in
$\D(M')$ extends to a diffeo of $\M$.  The extension is constructed by
specifying it to be the identity on $\M \backslash N$ where $N$ is an open
neighbourhood of $M'$ in $\M'$.  Within the neighbourhood we use the
isotopy sequence between $g$ and the identity in $D(M')$ to construct the
extension in the usual way.  Thus we have that the slide $s$ extends to
$\M$ and so $[s]$ acts trivially on $\Psi$.  Similarly the entire normal
subgroup $S$ generated by $[s]$ fixes $\Psi$ since any conjugate of
something isotopic to the identity or product of things isotopic to the
identity is also isotopic to the identity and
therefore extends to $^4\!M$.

So much for the slide.  We can also derive relations between the action of
the internal diffeos of one prime and the action of the internal diffeos of
the other on $\Psi$.  An internal diffeo of ``prime number one'', followed by
an internal diffeo of ``prime number two'' which undoes the twisting of the
U-tube caused by the first diffeo, will leave $\Psi$ invariant.  Now with
respect to our presentation, $G_1$ and $G_2$ are isomorphic via translation
$a:G_1\rightarrow G_2$, and 
(since $P$ is nonchiral) we can arrange that
there exists a second, ``mirror'' isomorphism,
$b:G_1 \rightarrow G_2$ which is given by reflection in the plane of
symmetry between the two cut-out polyhedra of $M$.  Then the element
$(g_1,g_2)\in G_1\times G_2$ extends if $g_2=b(g_1)$, and so $\Psi$ is
fixed by $(g_1, b(g_1))$.

\newsec{The Spin-Statistics Theorem}

Consider the space $\Lambda$ of all wave functions on the covering space
$\Q$.  Given a state vector in $\Lambda$, we can imagine decomposing it
into a superposition of components, each of which lives in a primary
subspace $\Lambda_\rho$ of $\Lambda$, where a primary subspace is the
direct sum of a number of copies of a single unitary irreducible
representation (UIR) $\rho$ of $G$.  The projection of a wave function onto
the primary subspace corresponding to a particular UIR $\rho$ is achieved
using the (un-normalized) operator:
\eqn\projec
{
   P_{\rho} =  \sum_{g \in G} \chi_\rho(g^{-1}) g
}
where $\chi_\rho$ is the character of the representation $\rho$.  Note
that $P_\rho$ commutes with every element of $G$.

We have seen that the normal subgroup $S$ of slides leaves our state $\Psi$
invariant.  Hence the only primary components that can occur in $\Psi$ are
those corresponding to UIR's in which the slides are represented trivially.
Proof: Let $\Pi$ project $\Lambda$ onto any {\it irreducible} component 
$\Lambda_{\rho{}0}\subseteq\Lambda$.  Since $\Pi\natural{}G$, we can write
its action consistently on the left.  Also let $\Psi_0=\Pi\Psi$.  Then
$\forall{}g\in{}G,\,\forall{}s\in{}S$ we have
$$
      \Psi_0 g s = \Psi_0 (g s g^{-1})g \ideq \Psi_0 s' g
$$
where $\s'\in S$ since $S$ is normal; and further
$$
  \Psi_0 s'=(\Pi\Psi)s'=\Pi(\Psi s')=\Pi\Psi=\Psi_0 ;
$$
hence $(\Psi_0 g)s=\Psi_0 g$.  But the $\Psi_0 g$ span the irreducible
subspace $\Lambda_{\rho{}0}$, whence $s$ must act as the identity on 
 $\Lambda_{\rho{}0}$.  Finally, since the choice of 
 $\Lambda_{\rho{}0}\subseteq\Lambda_\rho$ was arbitrary, it follows that
$s$ must act trivially on  $\Lambda_\rho$ itself.

The finite dimensional UIR's in which the slides are represented trivially
have been
classified in \refs{\rafsum}.  They are specified by: 
(i) a choice of an (unordered) pair $(\rho_1,\rho_2)$ of finite dimensional
    UIR's of the internal group of a single prime (say $G_1$), and 
(ii) a choice of sign for the exchange.  At the end of the previous section
we showed that certain elements $(g_1,g_2)\in G_1\times G_2$ act trivially
on $\Psi$.  From this we can also deduce a condition on $\rho_1$ and
$\rho_2$, namely that they are ``$CP$ conjugate'' representations, these
being defined by $\rho_1(g)=\rho_2(a^{-1}b(g))$.
If $\rho_1$ and $\rho_2$ are inequivalent UIR's then the quantum geons they
describe will be distinguishable particles, and no question of statistics
will arise.  If, on the other hand, $\rho_1$ and $\rho_2$ are equivalent,
then the geons will be identical, so let us now concentrate on those
particular primary subspaces.


Let us assume further that the representation $\rho_1$ that determines the
physical type of the geons is {\it  abelian}, in which case $\rho$ itself
is also 
abelian \refs{\rafsum}.  This means that $\rho$ represents every element of
the mapping class group $G$ by a pure number.  Now consider the
``component'' $\Psi_{\rho}:=\Psi{}P_\rho$ of $\Psi$ in the subspace
$\Lambda_\rho$.  We have, $\Psi_\rho{}F$ = $\Psi{}P_\rho{}F$ =
$\Psi{}F{}P_\rho$ = $\Psi{}P_\rho$ = $\Psi_\rho$.  Unless $\Psi_\rho$
vanishes, this means that $F$ acts in $\Lambda_\rho$ as the number $+1$.  In
particular, this implies that if $\Phi\in\Lambda_\rho$ is any wave function
in the subspace on which the loop $\Gamma$ can act (namely a wave function
peaked on a geometry describing two identically configured well-separated
geons), then $\Gamma$ acts trivially, or equivalently the exchange and
rotation {\it loops} act identically to each other.

Therefore, if a quantum sector carries abelian internal representations
$\rho_1=\rho_2$, and if $\Psi$ has support on that sector, then that
sector respects the spin-statistics correlation in the following sense.
For any state $\Psi$ in the sector on which the exchange and
$2\pi$-rotation loops act, both loops act identically.  We will call such
sectors spin-statistics respecting (and we note that all other sectors
which carry an abelian representation of the MCG are spin-statistics
violating).

Although this proof involves some subtle points, its main idea is simply
expressed.  Our choice of a particular presentation of the manifold $M$ has
the effect of {\it labeling} the geons, and wave functions 
$\Psi:\Q\to\Complexes$ can therefore be thought of as functions on a
configuration space of labeled particles.  Then (under appropriate
conditions, these providing the subtleties), the diffeomorphism $f_e$ just
represents {\it exchange of labels}, while $f^i_{2\pi}$  represents
$2\pi$ rotation of the geon  labeled $i$.  The relation $\Psi F =\Psi$ then
says that exchange of labels is equivalent to rotation of the first geon,
which is the spin statistics correlation. 

We are apparently unable to say anything about the quantum sectors
corresponding to non-abelian internal representations.  This might have
been expected, since geons carrying non-abelian representations of their
internal diffeomorphism groups possess non-geometrical internal states (the
phenomenon of ``quantum multiplicity''), and consequently, even when they
respect the spin-statistics correlation, one can construct states on which
the exchange loop $\Gamma$ acts as minus one.  For example, suppose the
geons are bosons.  Take the state in which geon A is in internal state
``up'' and geon B is in internal state ``down'' and superpose this with the
state in which A is down and B is up, with a relative minus sign.  Then the
exchange will take this state to minus itself, but the geons are the
epitome of boson-hood nevertheless.  Excluding a non-abelian
spin-statistics violating sector will thus require a stronger condition
than just the equality $\Psi F=\Psi$.  (In fact, even the question of which
of the words ``boson'' and ``fermion'' to attach to which sectors
can become confusing in some nonabelian cases.)

There are primes for which our result is a more complete spin-statistics
theorem than for others, because their internal diffeomorphism groups are
already abelian, and therefore the restriction to abelian representations
is no restriction at all.  These are the lens spaces $L(p,q)$ with $q^2 =
-1\ {\rm mod}\ p$ (the restriction on $p$ and $q$ is necessary and
sufficient for the lens space to be non-chiral).  Their internal group is
$Z_2$, the non-trivial element being a $\pi$ rotation, so they are
tensorial.  (A result due to Don Witt \refs{\witt} states that they are the
only non-chiral primes (except for the handle) with finite internal group.
As far as we know, it is an open question whether there exists a
non-chiral
prime whose internal group is infinite
abelian.)  So lens space geons, pair created via the cobordism $\M$, must
be bosons.

In using the operator \projec\ to decompose our U-tube engendered
wave-function $\Psi$ into primary components, we've been rather cavalier
about the fact that the group $G$ is infinite and discrete.  This causes
two main problems.  First, there is no reason for $P_\rho\Psi$ to be
normalizable, and it certainly cannot be normalizable in the most important
cases, where $\rho$ is finite dimensional.  
Second, infinite discrete groups commonly possess primary representations
of types II and III, and when this is the case for $G$, the operator
$P_\rho$ belonging to such primaries does not seem to be well-defined, even
formally (especially for type III).  Indeed, the decomposition into
irreducibles of type II and III representations is not unique, and a type
II or III primary can apparently not be associated naturally with any UIR
at all.

These problems arise partly because we have chosen to use covering space
quantization as a familiar setting in which to discuss spin and statistics.
In a frozen topology setting, one could solve the normalization problem by
treating each UIR as an inequivalent quantum theory (described in terms of
a vector bundle) and normalizing state vectors separately within each
sector.  The Type II and III representations could be avoided by
restricting only to finite dimensional UIR's.  Neither device is possible
here, since we have topology change.  However, we believe that our work can
(and should) be expressed solely in terms of spacetime histories.  In a SOH
formulation, normalization and restriction to finite dimensional
UIR's appear to present no special difficulties.  We discuss the SOH
further in the conclusions section.

\newsec{Conclusions}

In summary, a wave function $\Psi$ which is given by a functional integral
over geometries on a ``U-tube pair creation cobordism'' $\M$ has no support on
certain ``theta sectors'' of canonical quantum gravity, namely those
corresponding to spin-statistics violating abelian representations of the
MCG.  In particular, lens space geons $L(p,q)$ with $q^2 = -1$ mod $p$, pair
created via the cobordism $\M$, satisfy a spin-statistics correlation.  The
lens spaces are tensorial -- the $2\pi$-rotation of a lens space is trivial
-- so the result rules out the possibility that they are fermions.

We had restricted ourselves to orientable, non-handle geons but
we can generalize our calculation to include non-orientable non-handles. 
In this case, the condition of non-chirality is not 
meaningful and the U-tube is always a cobordism between 
$\R^3$ and $\R^3 \#P\#P$ when $P$ is non-orientable. The steps of our
calculation follow just as for the orientable case. 

How do we re-express our work in spacetime terms?  A sketch of the
fixed-topology case was given in \refs{\rafsum}.  In the SOH framework the
fundamental dynamical input is a rule attaching a quantum amplitude to each
pair of truncated histories which ``come together'' at some ``time''
\refs{\raffour,\unimod,\drexel}.  
Let us call such a pair a ``Schwinger history''
for short, and its underlying manifold a ``Schwinger manifold''.  In the
case of quantum gravity, a truncated history is a Lorentzian manifold with
final boundary\foot
{This final boundary corresponds to the spacelike slice $\M$ of
 the canonical formulation.}
(and possibly initial boundary depending on the physical
context), and the ``coming together'' means the identification or ``sewing
together'' of the final boundaries.  Now different ways of sewing are
possible, related to each other by large diffeomorphisms of the final
boundary.  In general such a re-identification may or may
not lead to a diffeomorphic Schwinger manifold, but it never will if we
restrict ourselves to product spacetimes of the form $\R\times\M$, i.e. if
we exclude topology change (and if we limit ourselves to diffeomorphisms
vanishing on any initial boundaries which may be present).  In this case,
the mapping class group $G$ of $\M$ acts freely and transitively (albeit
non-canonically) on the set of Schwinger manifolds.

Now, without disturbing the classical limit of the theory or the local
physics, we can multiply the amplitude of each Schwinger history by a
complex ``weight'' $w$ depending only on the topology of the underlying
manifold (and on the two initial metrics, if initial boundaries are
present).  Somewhat analogously to \refs{\cecile}, one can then argue that
consistency requires that these 
complex
weights transform under some unitary
representation of $G$, and that sets of weights belonging to disjoint
representations ``do not mix''.  The pure cases are then the UIR's, and we
arrive again at the conclusion that each distinct UIR of the mapping class
group yields an inequivalent version or ``sector'' of quantum gravity with
frozen topology.  Notice here that the weight function $w:G\to\Complexes$
need not (and in general will not) be square summable over $G$ (the trivial
UIR of $G$ corresponds to $w(g)\ideq{}1$ for example).  Thus there is no
apparent normalization problem in the SOH formulation.

Now let us bring in topology change and consider a Schwinger pair of U-tube
cobordisms.  Our spin-statistics result translates into the statement that
it would be inconsistent to try to use for the Schwinger manifolds one
obtains from the different attaching maps, a set of weights carrying an
abelian spin-statistics violating UIR of $G$.  (We would find that we were
trying to attach different weights to manifolds in the same diffeomorphism 
equivalence class.)  Such spin-statistics violating possibilities are thus
ruled out when one allows topology change.

One might contemplate enhancing the status of our result by strengthening
the restriction on UIR's of $G$ from finite dimensionality down to one
dimensionality (i.e. by admitting only abelian UIR's).  This might be
going too far, however, not only because there is no evident physical basis
for such a drastic restriction, but also because every prime three manifold
whose internal group is known to us lacks abelian spinorial
representations.  A restriction to abelian representations, therefore,
might rule out spinorial geons altogether, which would not be desirable.
However, in some sense of the word `most', the internal group remains
unknown for most prime three manifolds --- the mysterious and multitudinous
hyperbolic primes --- and they might include among them primes with abelian
spinorial representations.  We do not know how likely this is, but
physically it would seem hard to explain the appearance of spin-1/2 without
being able to trace it to some underlying ``hidden'' degrees of freedom
which, in turn, would be reflected in the quantum multiplicity associated
with nonabelian representations.  In this sense we can use physical
reasoning to ``predict'' something about 3-manifold topology: there should
be no prime whose MCG admits one-dimensional spinorial UIR's.

We see this work as an indication that there is a spin-statistics theorem
``trying to get out'' of a sum-over-histories formulation of quantum
gravity, and that it seems indeed to be intimately connected with the
process of pair-creation, as predicted. We do not think, however, that a
full spin-statistics theorem (including results for chiral geons and
``primordial'' geons) can be proved without extra input to the SOH rules,
such as that suggested in \refs{\raflect}.

\bigskip
\bigskip
\centerline{\bf Acknowledgements}
\par\nobreak
It is a pleasure to thank Lee Brekke, Andrew Chamblin, 
John Friedman, Nico Giulini, Bob
Gompf, Jorma Louko, Trevor Samols, and Don Witt for help and interesting
discussions. R.D.S. was supported in part by NSF Grant \# PHY-9307570 and
H.F.D. by the DOE and NASA grant NAGW-2381 at Fermilab, by NSF Grant \#
PHY-9008502 at Santa Barbara and by the U.~S.~Department
of Energy under Grant \# DE-FG03-92-ER40701 at Caltech. She also thanks
the Instituto de Ciencias Nucleares at the Universidad
Nacional Autonoma de Mexico 
for hospitality during the completion of this work.

\listrefs

\end